\makeatletter
\def\@cite#1#2{$^{\hbox{\scriptsize{#1\if@tempswa , #2\fi}}}$}
\makeatother
\documentstyle[12pt]{article}
\newcommand{\pp}{\vskip 0.12in}

\renewcommand{\baselinestretch}{1.9}
\begin{document}
\renewcommand{\baselinestretch}{1.9}
\vspace*{-0.4in}
\begin{center}
\begin{Large}

Phonon Transmission Rate, Fluctuations, and Localization in Random
Semiconductor
 Superlattices:  \\
Green's Function Approach\\

\end{Large}
\end{center}
\vskip 0.05in
\begin{center}
Norihiko Nishiguchi and Shin-ichiro Tamura\\
{\it Department of Engineering Science, Hokkaido University,
Sapporo 060, JAPAN }\\
Franco Nori\\
{\it Department of Physics, The University of Michigan,
Ann Arbor, MI 48109-1120}
\end{center}
\vskip 0.08in
\begin{abstract}

We analytically study phonon transmission and localization in random
superlattices by using a Green's function approach.
\ We derive expressions for the average transmission rate and
localization length, or Lyapunov exponent, in terms of the superlattice
structure factor.  This is done by
considering the backscattering of phonons,
due to the complex mass density fluctuations,
which incorporates all of the forward scattering processes.
\ These analytical results are applied to two types of
random superlattices and compared with numerical simulations
based on the transfer matrix method.
\ Our analytical results show excellent agreement with the numerical data.
\ A universal relation for the transmission fluctuations versus the average
transmission is derived explicitly, and independently confirmed by
numerical simulations.
\ The transient of the distribution of  transmission to the log-normal
distribution for the localized phonons is also studied.

\end{abstract}
\pp
\newpage

\section{INTRODUCTION }
\pp
The long-standing interest in Anderson localization has greatly increased
following the understanding that it is common to both electrons and classical
waves in disordered media\cite{CWL}.  For the localization of acoustic phonons,
an experimentally relevant quantity to study would be the transmission
rate. The well-known basic feature is that phonons will not
propagate through a medium with a large amount of randomness.  In particular,
in one dimension any disorder is strong enough to induce the exponential
localization of eigenmodes, {\it i.e.}, their amplitudes decay exponentially
along the medium with a rate of decay called the localization length. So far,
phonon transmission rates have been measured in synthetic multilayered
systems, or superlattices (SL's), with both periodic and quasiperiodic
order\cite{tam}. Similar experiments with randomly-layered systems
would provide important information on the localization properties of
high-frequency acoustic phonons.

In a previous paper\cite{tn90}, we studied phonon propagation through
 random SL's by means of the transfer matrix method.
The random SL's considered are the multilayered systems where two kinds of
basic blocks of materials (which may or may not have internal structure)
are stacked at random.
\ Each sample has its own realization of randomness in the order of
constituent layers, and the phonon transmission rate versus frequency shows a
fine
spiky structure specific to the particular realization of disorder present in
that given sample.  This fluctuating transmission rate can be considered to be
a fingerprint of the sample, like the reproducible conductance
fluctuations found in electronic mesoscopic transport.
\ The ensemble average of the transmission rate over possible configurations
of the constituent layers smears out the fine structure,
but still leaves
global features of transmission dips and peaks.
\ The dips arise from phonon localization due to the interference
among backscattered phonons\cite{CWL}.
The peaks arise from the resonance  occurring when
appropriate matching conditions are satisfied between the phonon wavelength
and the thickness of a basic layer.

A significant result of Ref.~3 is that there exists a remarkable
correlation between the ensemble-averaged reflection rate and squared
SL-structure factor calculated analytically.
\ The latter describes the sum of the phonon amplitudes reflected from SL
interfaces. The smaller contributions coming from multiple phonon reflections
have been neglected.   We find that at the frequencies where the maxima
(minima) of the structure factor are attained, the reflection rate exhibits
peaks (dips).

\pp
The purpose of the present study is to establish quantitatively
the relation between the phonon transmission rate and the SL-structure factor.
Also, we examine in some detail the localization characteristic of
phonons injected into random SL's.  Our study is based on the Green's
function method which has originally been applied to the electronic
conductivity in one-dimensional metals\cite{abrikosov}.

\ In section II, we model the random SL's and formulate the phonon
transmission rate in terms of  Green's functions.
\ In section III, we derive integral equations for the Green's functions by
introducing complex mass-density fluctuations which induce the backscattering
of phonons.
\ The solutions of these integral equations, satisfying appropriate boundary
conditions, are given in terms of the S-matrix associated with the problem.
In section IV, we show that the S-matrix elements can be written in terms of
the SL-structure factor, which describes the phonon amplitude backscattered
from the complex mass-density fluctuations.
\ The Born approximation is employed for the scattering and the explicit
expression is given for the phonon mean-free-path.
In section V, we derive an analytical expression for the average
transmission rate $ \langle T \rangle$, as a function of the system size
divided by the phonon mean-free-path.
\ We also apply this formula to two types of random SL's  and
compare the analytical results with the numerical simulations based on
the transfer matrix method.
In section VI, the Lyapunov exponent, or the inverse localization length, is
obtained by using a differential recursion relation.
\ The Lyapunov exponent is given by the reciprocal of the mean-free-path and
exhibits a good agreement with the numerical simulations.
\ The universality of the transmission fluctuation is discussed in
 section VII.  In section VIII,  we study the probability distribution of
the transmission rate and confirm that the log-normal distribution is
applicable to the localized phonons.
\ A summary and discussions are presented in the last section.

\section{Model }
In the present study, we consider the phonon transmission through two types of
random SL's with different structures in their constituent layers.  The first
class of SL's, which we refer to as single-layer SL's, consist of two kinds
of unit blocks as illustrated in Fig.1(a); that is, each unit block is
composed of a single material ($A$ or $B$) with a definite thickness
($d_A$ or $d_B$).
\ The second class of SL's, denoted as double-layer SL's, also consist of two
kinds of unit blocks but each one has an internal structure (or basis) as
shown in Fig.1(b).
\ The unit blocks of the latter class (specified by indices 1 and 2) are
composed of two different kinds of layers of materials $A$ and $B$ with
definite thicknesses ($d_{1,A}$ and $d_{1,B}$ in the first block and
$d_{2,A}$ and $d_{2,B}$ in the second block).
\ We assume that the order of the materials $A$ and $B$ is the same in these
building blocks of double-layer SL's.  Single-layer random SL's are constructed
by stacking these two kinds of single-layer unit blocks at random with
equal occurrence probabilities, {\em i.e.}, 0.5.  A random stacking of
the two types of double-layer unit-blocks creates a double-layer random SL.
\ It should be noted that phonons cannot recognize the existence of the
interfaces between consecutive $AA$ and $BB$ layers which occur very often
in the single-layer SL's.  So, it is convenient to define a ``segment" which
stands for any part of a single-layer SL consisting of a consecutive series of
the same kind of blocks.  In short, a segment is a part of a SL consisting of a
{\it single\/} material $A$ or $B$.   In any double-layer SL, however,
different materials always meet at any interface.  Hence, for the double-layer
SL the segment is only a part of a unit block.  Hereafter, we call a segment
consisting of material $A\ (B)$ simply as an $A$ (a $B$) segment.

\pp
If the SL's considered are composed of two kinds of materials, $A$ and $B$,
 they can be modelled as an alternating sequence of segments of materials
$A$ and $B$ as described above.
The thicknesses of the segments, denoted by $D_A$ and $D_B$,
are distributed according to the prescribed structures
of the basic building blocks
and the probability for the successive stacking of the same kind of material.
\ For simplicity, $A$ and $B$ are assumed to have the same stiffness
constant  $\mu$  and are distinguished by their mass densities $\rho_A$  and
$\rho_B$.  We also assume that a random  SL is sandwiched between a substrate
and a detector composed of material $A$.

Now, we consider the situation where an acoustic wave excited in the substrate
propagates through a random SL and  is observed at the detector on top of the
SL.  We assume the propagation to be perpendicular to the interfaces of the
building blocks.  No mode conversion among phonons at the interfaces is
considered.
This is valid if the interfaces have a mirror symmetry.
\ As the randomness is due to the stacking
order of blocks  and is present only along the   propagating direction
({\em x}-direction),  the system we discuss is essentially one-dimensional.

\pp
\ The basic wave equation governing the motion of the displacement
$u(x,\omega )$ at an angular frequency $\omega $ is
\begin{equation}
(  \rho (x) \omega^2  + \mu \partial_x^2) u(x,\omega) = 0,
\label{eq.of.u}
\end{equation}
\noindent
where the mass density $\rho(x)$ takes on the value of either $\rho_A$ or
$\rho_B$ depending on the position $x$ $(0 < x < L)$ in the SL of length $L$.
Also, $\rho(x)$ is taken to be $\rho_A$ in the substrate
$( x < 0 )$  and detector $( x > L )$.  We define
$\delta\bar\rho=\rho_B-\rho_A$.
For convenience, we assume $ \rho_A < \rho_B$ but the results
are also valid for $\rho_A > \rho_B$.

We formulate the transmission in terms of the retarded and advanced Green's
functions associated with Eq.(\ref{eq.of.u}):
\begin{equation}
( \rho (x) \omega_\pm^2 + \mu\partial_x^2) G_{\omega \pm}(x,x') = \delta
(x-x'),
\label{eq.of.G}
\end{equation}
\noindent
where  $\omega_\pm = \omega \pm i \delta $  and  $\delta$ is
an infinitesimal positive number.
We define the transmission rate by the ratio of the acoustic  Poynting
vector in a random SL to that in a homogeneous system with the mass density
$\rho_A$.  The unperturbed Green's functions satisfy the equation:
\begin{equation}
( \rho_A \, \omega_{\pm }^2 + \mu \partial _{x }^2 ) G_{\omega \pm}^0 (x,x') =
\delta (x-x')
\label{eq.of.G0}
\end{equation}
\noindent
and the solutions are
\begin{equation}
G_{\omega\pm}^0(x,x') = \mp \frac{i}{2 \rho_Ac_A\omega}
      e^{  \pm i \omega |x-x'| /c_A },
\label{G0}
\end{equation}
where $c_A$ is the sound velocity in the $A$ material which is given by
\begin{equation}
c_A = \sqrt{\mu/\rho_A} \; ,
\end{equation}
and $c_B$ is the sound velocity in the $B$ material,
\begin{equation}
c_B = \sqrt{\mu/\rho_B} \; .
\end{equation}
Thus, the transmission rate $T(L,\omega )$ is written with $G_{\omega\pm}$ as
\begin{equation}
T(L,\omega) = (  2\, \rho_A \, c_A \, \omega)^2 G_{\omega+}(L,0)\
G_{\omega-}(L,0).
\label{def.of.T}
\end{equation}
\noindent

\section{One-particle Green's function}

The mass density fluctuation $\delta \rho (x) = \rho (x) - \rho_A$ present
in the SL causes the forward and backward scatterings of phonons incident
upon the SL from the homogeneous substrate.
A key observation is that the effect of the forward scattering processes can be
exactly summed up.
The Green's function $\tilde{G}^0$ incorporating all of the forward scattering
processes is related to the unperturbed Green's function
$G^0$ as
\begin{equation}
\tilde{G}_{\omega+}^0 (x,x')\ = \ G_{\omega+}^0 (x,x') \
e ^{i\, sign(x-x')\left[\Phi (x) \ -\ \Phi(x')\right]},
\label{G0B}
\end{equation}
where the phase  $\Phi$  is defined 
by
\begin{equation}
\Phi (x) \ = \ \frac{\omega}{2\rho_A c_A} \int_{0}^x \delta \rho (y) dy.
\label{Phi}
\end{equation}
Now, considering  only backscattering processes,
we can express the solution of Eq.(\ref{eq.of.G})
in terms of the phase-modulated  Green's functions,
 $\tilde{G}_{\omega+}^0$'s.
For an even number of backscattering processes, and for $x>x'$,
the retarded Green's function  satisfies
\begin{eqnarray}
G_{\omega+}(x,x')\ & = & \ \tilde{G}_{\omega+}^0(x,x')\ +
\int_{0}^{L} \int_{0}^{L} \
\tilde{G}_{\omega+}^0(x,x_1) \left( - \delta \rho (x_1)
\omega^2\right) \tilde{G}_{\omega+}^0(x_1,x_2)(-\delta \rho(x_2)
\omega^2) \nonumber \\
&& \times
G_{\omega+}(x_2,x') \theta(x_2-x_1)
 \theta(x-x_1) \theta(x_2-x')  dx_1 dx_2.
\label{GE1}
\end{eqnarray}
In order to express Eq.(\ref{GE1}) with the bare Green's function
$G_{\omega+}^0$, we multiply both sides of Eq.(\ref{GE1}) by
$\exp[ - i (\Phi  (x) - \Phi (x'))]$ and introduce
\begin{equation}
\hat{G}_{\omega+}(x,x')=G_{\omega+}(x,x') e^{ -i \, [\Phi(x)-\Phi(x')]}.
\label{Gbare}
\end{equation}
Thus, we obtain
\begin{eqnarray}
\hat{G}_{\omega+}(x,x')\ &=& G_{\omega+}^0(x,x')\ +
\int_{0}^{L} \int_{0}^{L}
\  G_{\omega+}^0(x,x_1) \left( - \zeta(x_1) \omega^2\right)
  G_{\omega+}^0(x_1,x_2)\left(-\zeta^\ast(x_2)
\omega^2 \right) \nonumber  \\
&& \times \hat{G}_{\omega+}(x_2,x')
\theta(x_2-x_1) \theta(x-x_1)\theta(x_2-x') dx_1 dx_2 ,
\label{GE2}
\end{eqnarray}
where $\zeta(x)$ denotes the complex mass density fluctuation defined by
\begin{equation}
\zeta (x) \ = \ \delta \rho(x) e^{- 2 i\Phi(x)}.
\label{zeta}
\end{equation}

As the complex mass density fluctuations,
$\zeta^\ast$  and $\zeta$, transfer a phonon with a wave number
$k (= \omega / c_A) $  to $ -k$  and vice versa, it is convenient to adopt
a $2\times2$ matrix representation for the Green's functions,
$[g_{\omega +}]_{ij}$,
 where the indices $i, j = 1 \ {\rm and} \  2$ correspond to the phonon states
with
 $k$ and $ -k$, respectively.
Defining the diagonal matrix elements for the unperturbed Green's function,
$g^0_{\omega +}$, as
\begin{eqnarray}
[g^0_{\omega +}(x,x')]_{11} & = &G^0_{\omega +}(x,x') \, \theta ( x - x' )
\nonumber \\
&=& - \frac{i}{2\rho_A c_A\omega }
\ \theta (x-x') e^{ i \omega (x-x')/{c_A}}
\label{g0}
\end{eqnarray}
and
\begin{equation}
[g^0_{\omega +}]_{22}(x,x') = G^0_{\omega +}(x,x') \, \theta ( x' - x ),
\label{gbare22}
\end{equation}
we re-express Eq.(\ref{GE2}) in terms of the matrix elements of
$g_{\omega +}$ and $g_{\omega +}^0$ as
\begin{eqnarray}
[g_{\omega +}(x,x')]_{11}\ &=& [g^0_{\omega +}(x,x')]_{11}\ +
\int_{0}^{L} \int_{0}^{L}
[g^0_{\omega +}(x,x_1)]_{11} \left( - \zeta(x_1) \omega^2\right)
  [g^0_{\omega +}(x_1,x_2)]_{22} \nonumber  \\
&& \times \left(-\zeta^\ast(x_2)
\omega^2 \right)[g_{\omega +}(x_2,x')]_{11} dx_1 dx_2 ,
\label{gE}
\end{eqnarray}
where $[g_{\omega +}]_{11}$ is identical to $\hat{G}_{\omega +}$ defined
by Eq.(\ref{Gbare}) and the
indices indicate that it begins and terminates with $[g^0_{\omega +}]_{11}$.
The other diagonal element $[g_{\omega +}]_{22}$ is obtained from Eq.(\ref{gE})
by interchanging $\zeta$ and $\zeta^\ast$ and the indices $1$ and $2$.

We can treat the case of odd numbers of backscattering events in the same way.
The off-diagonal elements $[g_{\omega +}]_{21}$ defined by
\begin{equation}
[g_{\omega +} (x,x')]_{21}=G_{\omega+}(x,x')\exp[ \ i \ (\Phi (x) + \Phi(x'))]
,
\label{Gbaro1}
\end{equation}
can be expanded in terms of
the complex mass  density fluctuations $\zeta$  and $\zeta^\ast$   as follows:
\begin{eqnarray}
 [g_{\omega +}(x,x')]_{21}  &= &
\int_{0}^{L} [g^0_{\omega +}(x,x_1)]_{22}  \left( - \zeta^\ast(x_1)\omega^2
\right)
[g^0_{\omega +}(x_1,x')]_{11}  dx_1 \nonumber \\
&&  +\int_{0}^{L} \int_{0}^{L} [g_{\omega +} (x,x_2) ]_{21}
 \left(-\zeta(x_2) \omega ^2\right)
[g^0_{\omega +} (x_2,x_1) ]_{22} \nonumber \\
&& \times  (- \zeta^{\ast}(x_1) \omega^2) \
[g^0_{\omega +}(x_1,x')]_{11}   dx_1 \ dx_2.
\label{GO1}
\end{eqnarray}
\noindent
The expression for the element $[g_{\omega +}]_{12}$ defined by
\begin{equation}
[g_{\omega +}(x,x')]_{12} =G_{\omega+}(x,x')\exp[- \ i \ (\Phi (x) + \Phi(x'))]
\label{Gbaro2}
\end{equation}
is obtained from  Eq.(\ref{GO1})
by interchanging the $\zeta$ and $\zeta^\ast$,   and also
the indices $1$ and $2$.
\
Now, we can rearrange the equations for $g_{\omega+}$, by employing  Pauli's
spin matrices $ \sigma_3$ and
$\sigma^{(\pm)}\ = \ \sigma_1 \pm i \sigma_2 \, $,  to give
\begin{equation}
\left\{ 2\rho_A \omega \left[
        i c_A \sigma_3 \partial_x + \omega_{+}
\right] -
 \frac{\omega^2}{2} \left[
       \sigma^{(+)} \zeta(x) + \sigma^{(-)} \zeta^\ast(x)
\right]
\right\}
 g_{\omega+}(x,x')  =  \delta (x-x').
\label{eq.of.g}
\end{equation}
\noindent
To solve Eq.(\ref{eq.of.g}), we follow the mathematical method developed
by Abrikosov and Ryzhkin for treating the electronic conductivity problem
in one-dimension\cite{abrikosov}.

First of all, we introduce the $2 \times 2 \ S$-matrix
which obeys the following equation :
\begin{equation}
\left\{
2 \rho_A \omega \left[ i c_A \sigma_3 \partial_x + \omega \right]
 - \frac{\omega^2}{2} \left[ \sigma^{(+)} \zeta(x)+\sigma^{(-)} \zeta^\ast(x)
\right]  \right\}
S_{\omega}(x,x')=0.
\label{eq.of.S}
\end{equation}
As the terms with complex mass density fluctuations $\zeta$ and $\zeta^\ast$ do
not commute with each other,
the $S$-matrix can be expressed with the position-ordering operator $T_x$
as
\begin{equation}
S_\omega(x,x') = T_x \exp  \left\{ i \frac{\omega}{c_A} \int_{x'}^{x}
 \Bigl\{ \sigma_3 - \frac{1}{4 \rho_A} \left[\sigma^{(+)}
 \zeta(x_1) - \sigma^{(-)}\zeta^\ast(x_1)\right] \Bigr\}dx_1 \right\}.
\label{S}
\end{equation}
\noindent
The operator $T_x$  rearranges a product of position-dependent
operators so that the operators must be placed from right to left
in the order of increasing $x$.  For $x>x_1>x'$, the retarded Green's function
$g_{\omega +} (x, x')$ is related to
$g_{\omega +} (x_1 , x')$ by the expression
 \begin{equation}
g_{\omega + }(x,x') = S_\omega(x,x_1) g_{\omega+}(x_1,x').
\label{gS1}
\end{equation}
The above relation also holds for 
$x'>x>x_1 $.
However, for $x>x'>x_1  $,  the relation between $g_{\omega +} (x, x')$
and $g_{\omega +} (x_1 , x')$ is not given by Eq.(\ref{gS1}) because
$g_{\omega +} (x, x')$ has an additional term arising from the
$\delta$-function in
Eq.(\ref{eq.of.g}). After a bit of algebra, we obtain
\begin{equation}
g_{\omega + }(x,x') = S_\omega(x,x_1) g_{\omega+}(x_1,x')-
\frac{i}{2\rho_A c_A \omega} S_\omega(x,x') \sigma_3.
\label{gS2}
\end{equation}
Equations (\ref{eq.of.g}), (\ref{gS1}) and (\ref{gS2})  hold for
the advanced Green's function as well as for the retarded Green's
function. 

Now, we impose the boundary conditions satisfied by the retarded Green's
function on Eqs.(\ref{gS1}) and (\ref{gS2}).
Noting that the perturbation series of  Eq.(\ref{gE}) for
$[g_{\omega +}(x,x')]_{11}$ begin and terminate with
$[g^0_{\omega +}]_{11}$
and that all terms in the iterative solution
$[g^0_{\omega +}(x, x')]_{11}$ vanish in the
limit of $x' \rightarrow \infty $ or  $x \rightarrow - \infty$,
the boundary condition for $[g_{\omega +}(x, x')]_{11}$ becomes
\begin{equation}
[g_{\omega +}(x, x')]_{11} \rightarrow 0 \hspace{0.1in} {\rm at} \hspace{0.1in}
  x  \rightarrow - \infty
      \hspace{0.1in}  {\rm or} \hspace{0.1in} x' \rightarrow \ \infty.
\label{bc1}
\end{equation}
 In the same way,  we obtain the following boundary
conditions for the other elements of the retarded Green's function:
\begin{eqnarray}
\begin{array}{ccccc}
[g_{\omega +}(x, x')]_{22}  \rightarrow   0  &  {\rm at} &
          x \rightarrow  \ \infty  &  {\rm or} &  x' \rightarrow - \infty,
\label{bc2}
\end{array}
\end{eqnarray}
\begin{eqnarray}
\begin{array}{ccccc}
[g_{\omega +}(x, x')]_{12}  \rightarrow 0  &  {\rm at} &
   x \rightarrow - \infty  &  {\rm or} &   x' \rightarrow - \infty,
\label{bc3}
\end{array}
\end{eqnarray}
\begin{eqnarray}
\begin{array}{ccccc}
[g_{\omega +}(x, x')]_{21}  \rightarrow 0 &  {\rm at} &
   x \rightarrow \ \infty   &  {\rm or} &      x'  \rightarrow  \ \infty.
\label{bc4}
\end{array}
\end{eqnarray}
Taking the limit of $x_1 \rightarrow - \infty$ in Eq.(\ref{gS2}),
we obtain from the boundary conditions (\ref{bc1}) to
(\ref{bc4}),
\begin{equation}
[g_{\omega+}(x,x')]_{ \alpha \beta}  =
   [S_{\omega}(x, - \infty)]_{\alpha 2} [g_{\omega +}(- \infty,x ')]_{2 \beta}
   - \frac{i}{2\rho_A c_A \omega} \left[ S_{\omega}(x,x') \sigma_3
   \right]_{\alpha \beta} \theta (x-x').
\label{gS3}
\end{equation}
If we take the limit of $x\rightarrow \infty$ and put $\alpha=2$, the left
hand side of Eq.(\ref{gS3}) vanishes.
Thus, we find
\begin{equation}
[g_{\omega+}(-\infty,x')]_{2\beta} = \frac{i}{2\rho_A c_A \omega} \
\frac{1}{[S_{\omega}(\infty,-\infty)]_{22}}
\left[S_\omega(\infty,x') \sigma_3\right]_{2\beta}.
\label{gS4}
\end{equation}
Substituting Eq.(\ref{gS4})  into Eq.(\ref{gS3}), we can express the retarded
Green's function as follows:
\begin{equation}
[g_{\omega+}(x,x')]_{\alpha\beta}=\frac{i}{2\rho_A c_A \omega}
        \left\{
   \frac
   {[S_{\omega}(x,-\infty)]_{\alpha 2}   \left[S_\omega(\infty,x') \sigma_3
   \right]_{2 \beta}}
   {[S_{\omega}(\infty,-\infty)]_{22} }
-\left[S_\omega(x,x')\sigma_3\right]_{\alpha\beta }  \theta(x-x') \right\}.
\label{gR}
\end{equation}
Similarly, the advanced Green's function can be expressed as
\begin{equation}
[g_{\omega-}(x,x')]_{\alpha\beta}=\frac{i}{2\rho_A c_A \omega}
\left\{ \frac
{[S_{\omega}(x,-\infty)]_{\alpha 1}
\left[S_\omega(\infty,x') \sigma_3 \right]_{1 \beta}}
{[S_{\omega}(\infty,-\infty)]_{11} }
-\left[S_\omega(x,x')\sigma_3\right]_{\alpha\beta }  \theta(x-x') \right\}.
\label{gA}
\end{equation}
\noindent
Equations (\ref{gR}) and (\ref{gA}) are used  to derive the  transmission rate.

To proceed further, we introduce  an interaction representation of the
S-matrix with respect to $\omega$
\begin{equation}
S_\omega(x,x') = e^{ i \omega x \sigma_3 /c_A} S(x,x')
    e^{- i \omega x' \sigma_3/c_A}.
\label{SII}
\end{equation}
Explicitly,
\begin{equation}
S(x,x') = T_x \exp  \left\{ - i \frac{\omega}{4\rho_A c_A} \int_{x'}^{x}
 \left[\sigma^{(+)}
 \zeta_{\omega}(x_1) - \sigma^{(-)}\zeta_{\omega}^\ast(x_1)\right]
dx_1 \right\},
\label{SI}
\end{equation}
where
\begin{eqnarray}
\frac12 \sigma^{(+)} \zeta_\omega(x)\ & =
    & e^{- i \omega x \sigma_3/c_A} \frac12 \sigma^{(+)} \zeta(x)
e^{ i \omega x \sigma_3/c_A} \nonumber \\
\vspace{0.3in}
&=&\ \frac12 \sigma^{(+)} \zeta(x) e^{- 2 i \omega x/c_A}.
\label{sz}
\end{eqnarray}
It should be noted here that the S-matrix $S(\infty, - \infty)$ in the
interaction representation is identical to  $S(L, 0)$, or
\begin{equation}
S(\infty, - \infty) = S(L,0),
\label{SL0}
\end{equation}
since $\zeta(x)$ and $\zeta^\ast(x)$ vanish outside
({\it i.e.}, $x > L \ {\rm and } \ x < 0$) the SL.

Next, we divide the system into a set of slabs with thickness
$\Delta$ and perform the integral of Eq.(\ref{SI}). To do this, we define
$\kappa_j$ by
\begin{equation}
\kappa_j \ = \ - i \frac{\omega}{2\rho_A c_A} \int_{x_j}^{x_{j+1}}
\zeta_\omega(y) \ dy,
\label{kappa}
\end{equation}
where $x_j \left(=j \Delta \right)$ denotes the position of interface between
 the $j$th and
$(j+1)$th slabs.
In terms of $\kappa_j$ and its complex conjugate
the integral in Eq.(\ref{SI}) is replaced with a sum  and the S-matrix
becomes a product of the factors containing $\kappa_j$ and $\kappa^\ast_j$
at different slabs.
Keeping terms to second order in $\kappa_j$ and
$\kappa_j^\ast$, we obtain
\begin{equation}
S(x,x') \ = \ \prod_{j =x'/\Delta}^{x/\Delta} \
\left(
 \begin{array}{cc}
       1+\frac12 |\kappa_j|^2   &        \kappa_j \\
        \kappa_j^{\ast}                      & 1+\frac12 |\kappa_j|^2
  \end{array}
\right).
\label{S.mat}
\end{equation}
We note here that the determinant of the $S$-matrix is unity to order
$|\kappa|^2$. \

\pp
\section{Backscattering Rate and SL Structure Factor}

The quantities $\{\kappa_j\}$ constituting the S-matrix are a set of
fundamental random variables which vary from sample to sample.
The statistical properties of
$\{\kappa_j\}$ are expected to reflect the structures
of the basic building blocks of random SL's.
Equation (\ref{kappa}) can be rewritten in terms of the mass density
fluctuation
$\delta\rho(x)$  and sound velocities in the constituent materials $A$ and $B$.
Substituting Eqs.(\ref{zeta}) and (\ref{sz}) into
(\ref{kappa}), we have
\begin{equation}
\kappa_j  = - i \frac{\omega}{2\rho_A c_A}
\int_{x_j}^{x_{j+1}} dy \delta \rho(y)
\exp
       \left[
      -  i   \int_{0}^{y} \frac {2 \omega} {c(y')}  dy'
\right] .
\label{kappa.int}
\end{equation}
The sound velocity $c(x)$ has value $c_A$  in $A$ and $c_B$  in $B$.
In deriving Eq.(\ref{kappa.int}),
we have used
\begin{equation}
c_B \simeq  c_A \left( 1+ \frac{\delta \bar \rho}{2 \rho_A} \right)^{-1}.
\label{s.vel}
\end{equation}

Next we define the average of variables $\left\{\kappa_j\right\}$
in a random SL  by
\begin{equation}
\overline{\kappa_j}  = \ \frac{\Delta}{L} \ \sum_{j=0}^{L/\Delta} \ \kappa_j.
\label{kappa.sum}
\end{equation}
Putting Eq.(\ref{kappa.int}) into (\ref{kappa.sum}),
we can relate $\overline{\kappa_j}$ to the structure factor $S_{SL}(L,\omega)$
of
a random SL as follows:
\begin{eqnarray}
\overline{\kappa_j} & = & \frac\Delta{L}  \left( \frac{- i \omega}
{2\rho_A  c_A} \right)
\int_0^L dy\delta\rho(y)\exp\left[- i\int_0^{y}
\frac{2\omega}{c(y')}dy'\right]   \nonumber \\
& & \nonumber \\
& = & \frac\Delta{L}  \frac{ \delta\bar{\rho} \, c_B} {4\rho_A \, c_A}
 S_{SL}(L,\omega),
\label{k.to.stf}
\end{eqnarray}
where
\begin{equation}
S_{SL}(L,\omega)  \equiv  \left(\frac{-2i\omega}{c_B\delta\bar{\rho}} \right)
\int_0^L dy\delta\rho(y)\exp\left[- i\int_0^{y}
\frac{2\omega}{c(y')}dy'\right].
\end{equation}
By performing the integral for a random SL which begins
with segment $B$, the structure factor $S_{SL}(L,\omega )$ is
written  as
 the superposition of amplitudes of phonons reflected once from each
interface
\begin{equation}
S_{SL}(L,\omega)\ = \ \sum_{j=1}^M (1-e^{- i \theta_{2j-1}})
     \exp\left[- i \sum_{m=0}^{2j-2} \theta_m \right],
\label{str.fct}
\end{equation}
where $ \theta_m=2\omega D_m/c_m$ (put $\theta_0\equiv 0$),
 $D_m$ is the thickness of the {\em m}th segment, $c_m$ is the sound
velocity, and $M$ is the number of $B$ segments.
The expression of the structure factor for a random SL
which begins with segment $A$ is similarly given by
\begin{equation}
S_{SL}(L,\omega)\ =  \sum_{j=1}^M (1-e^{- i \theta_{2j}})
     \exp\left[- i \sum_{m=0}^{2j-1} \theta_m \right].
\end{equation}
 The size of the random SL, $L$, is the sum of
the thicknesses of its constituent segments
\begin{equation}
   L = \sum_{m=1}^N D_m,
\label{LN}
\end{equation}
where $N$ is the total number of segments in the SL.

Now we consider the ensemble average $\langle \overline{\kappa_j} \rangle$
of  $\overline{\kappa_j}$ ($\propto S_{SL}$) over
possible  realizations of random sequences of the segments.
 However, $\langle \overline{\kappa_j} \rangle$  vanishes because
 $S_{SL}(L,\omega )$ is a complex-valued random variable.
\ The ensemble average of the product $\kappa_j \kappa_k^{\ast}$
also vanishes for $j\ne k$ by the same reason.
The non-vanishing term $\langle \overline{ |\kappa_j|^2} \rangle$   can  be
given in terms of the structure factor $S_{SL}(L,\omega )$ as follows:
\begin{eqnarray}
\langle \overline{ |\kappa_j|^2} \rangle \ & = & \
              \frac{L}{\Delta} \langle |\overline{\kappa_j }|^2 \rangle
\nonumber \\
& = &
 \  \frac{\Delta}{L}  R^2\ \langle |S_{SL}(L,\omega)|^2 \rangle.
\label{av.kk1}
\end{eqnarray}
Here we note that, to lowest order in the mass density difference
$\delta\bar{\rho}$,
 the prefactor $ \delta\bar{\rho} \, c_B/4\rho_A \, c_A$ in Eq.(\ref{k.to.stf})
is equivalent to  the reflection coefficient $R_{BA}$ of an acoustic wave
incident on the interface from the material $B$ to the material $A$,
\begin{eqnarray}
R_{BA} &= & \frac{Z_B - Z_A}{Z_B+Z_A}  \nonumber \\
    &= & - R_{AB}  \equiv R \ .
\end{eqnarray}
Here, $Z_A$ and $Z_B$ are the acoustic impedances and are defined
by $Z_A=\rho_A c_A$ and $Z_B=\rho_B c_B$.

The quantity $\langle \overline{ |\kappa_j|^2} \rangle$ is related to the
phonon backscattering rate $\Gamma$ caused by the complex mass density
fluctuation $\zeta(x)$.  This can be seen by expressing
$\langle \overline{ |\kappa_j|^2} \rangle$  in terms of $\zeta(x)$.
{}From Eqs.(\ref{sz}) and (\ref{kappa}),
$\langle \overline{ |\kappa_j|^2} \rangle$
can be rewritten as
\begin{equation}
\langle \overline{ |\kappa_j|^2} \rangle
 \ = \ \Delta  \frac{\omega^2}{4\rho_A^2 c_A^2 L}
\left\langle \left| \int_0^L \zeta(x) e^{-2 i \omega x /c_A} dx
\right|^2 \right\rangle.
\label{av.kk2}
\end{equation}
The backscattering rate $\Gamma$ is calculated  in the Born approximation as
\begin{equation}
\Gamma(L,\omega)\ = \
\frac{\omega^2}{4\rho_A^2 c_A L}
\left\langle \left| \int_0^L \zeta(x) e^{-2 i \omega x /c_A} dx
\right|^2 \right\rangle.
\label{Gamma}
\end{equation}
Comparing Eqs.(\ref{av.kk1}), (\ref{av.kk2}) and (\ref{Gamma}),
the mean-free-path
$\ell$ limited by backscattering is
\begin{equation}
\ell(L,\omega)^{-1} = \frac{\Gamma(L,\omega)}{c_A} = R^2
\frac{\langle|S_{SL}(L,\omega)|^2\rangle}{L},
\label{ell}
\end{equation}
so we can express $\langle \overline{ |\kappa_j|^2} \rangle$ in terms of $\ell$
as
\begin{equation}
\langle \overline{ |\kappa_j|^2} \rangle  \ = \
\frac{\Delta}{\ell(L,\omega)}.
\label{av.kk3}
\end{equation}

In the rest of this section, we analytically calculate
$\langle|S_{SL}(L,\omega)|^2\rangle$ for  the random SL's.
For an odd $N$, the 
first and last segments of a
random SL are the same.
For an even $N$, a random SL which begins with an $A\ (B)$ segment terminates
with a $B\ (A)$ segment.  For a double-layer SL, $N$ is always even and it
begins with an $A$ segment and terminates with a $B$ segment.
If an end segment of a SL is an $A$ segment, there is no phonon  reflection
at the boundary between  the random SL and the adjacent substrate or detector
(assumed to be made of $A$ material).
In this case, we regard the end segment   as a part of
the substrate or detector.
Hence, we may consider the random SL's  whose end segments are $B$ and
take $N=2M-1$.
Thus the length of the random SL becomes
$L = (\langle D_A\rangle+\langle D_B\rangle) M$ for large $M$.
Here $\langle D_A\rangle$ and $\langle D_B\rangle$ denote the averaged
thicknesses of the $A$ and $B$ segments, respectively.

By using Eq.(\ref{str.fct}), the squared SL structure factor becomes
\begin{eqnarray}
\left| S_{SL}(L,\omega)\right|^2 \ & = &\
       \sum_{n=1}^M \left( 1- e^{-i \theta_{2n-1}} \right)  \nonumber \\
  \; \; \; \;    & & + \sum_{m=2}^M \sum_{n=1}^{m-1}
            \left(1-e^{- i \theta_{2m-1}}\right)
                  \left(e^{- i \theta_{2n-1}}-1\right)
                     \exp \left[- i \sum_{j=2n}^{2m-2}\theta_j\right]     +  \;
c.c.
\label{SS}
\end{eqnarray}
Since  there is no correlation between the lengths of adjacent segments,
$\theta_j$'s associated with an $A$ segment
$ (\theta_j=\theta_A \equiv 2 \, \omega \, D_A /\, c_A)$ and $B$ segment
$ (\theta_j=\theta_B \equiv 2 \, \omega \, D_B /\, c_B)$  are independent
variables.
  From these considerations, the ensemble average of the
squared  SL structure factor per segment,  $I_s$, defined by
$I_s = \langle|S_{SL}(L,\omega)|^2\rangle / N$,  becomes\cite{tn90}
\begin{equation}
I_s(L,\omega)\ = Re \left[
\ \frac{(1-\epsilon_A)(1-\epsilon_B)}{1-\epsilon_A \epsilon_B}\right] +
 \frac{\langle D_A\rangle+\langle D_B\rangle}{L} Re\left[
\left(\frac{1-\epsilon_B}{1-\epsilon_A \epsilon_B} \right)^2 \ \epsilon_A \
\right],
\label{Is}
\end{equation}
where             $\epsilon_A = \langle e^{- i \theta_A} \rangle $
and
         $\epsilon_B = \langle e^{- i \theta_B} \rangle $.
Equation (\ref{Is}) gives the backscattering rate in the random SL's
(through Eq.(\ref{ell})).
The second term of Eq.(\ref{Is})  depends on the system size $L$ and
vanishes  in the limit $L \rightarrow \infty$.

Next we consider the case for $\epsilon_A=1$ and $\epsilon_B\neq1$.
This  is one of the resonance conditions under which phonons can be transmitted
perfectly through the SL.
This condition is satisfied, for instance, when twice the thickness of
 $A$ layer or $A$ block  is an integer-multiple of the wavelength of phonons.
In this case,  the SL can be effectively regarded as a bulk layer of material
$B$.  However, the phonon  reflections  at both ends of the SL still exist.
Under this condition,  only the second term of Eq.(\ref{Is}) survives.
Thus, we can attribute the second
term to the scattering at the  SL-substrate and the SL-detector
 boundaries, which is important at frequencies close to the resonance.
However, for $\epsilon_A\neq1$ and $\epsilon_B=1$, {\it i.e.},
at the resonance for $B$ segments, both terms in Eq.(\ref{Is}) vanish.
This asymmetry originates from the fact
that the SL is sandwiched between materials of type A.

Based on the  above considerations, we may separate the mean-free-path $\ell$
into two parts: one limited by scattering within the SL,
$\ \ell_{SL}$, and the other limited by boundary scattering, $\ell_b$.  Thus,
\begin{equation}
\ell^{-1} = \ell_{SL}^{-1} + \ell_b^{-1},
\label{ell.s}
\end{equation}
where
\begin{equation}
\ell_{SL}(\omega)^{-1}= \frac{2 R^2}{\langle D_{A}\rangle +
 \langle D_{B}\rangle}
 Re \left[ \frac{(1-\epsilon_{A})(1-\epsilon_{B})}{1-\epsilon_{A}\epsilon_{B}}
\right],
\label{ell.SL}
\end{equation}
and
\begin{equation}
\ell_b(L,\omega)^{-1} = \frac{2 R^2}{L}
 Re \left[ \left(\frac{1-\epsilon_{B}}{1-\epsilon_{A}\epsilon_{B}}\right)^2
\epsilon_A\right].
\label{ell.b}
\end{equation}
\section{TRANSMISSION RATE}
Noting that $g_{\omega\pm}$'s differ from $G_{\omega\pm}$'s
 by phase factors only
[see Eqs.(\ref{Gbare}), (\ref{gE}), (\ref{Gbaro1}) and (\ref{Gbaro2})],
we  can re-express the ensemble averaged transmission rate $\langle T \rangle$
 in terms of the Green's functions $g_{\omega\pm}(x,x')$ as
\begin{equation}
\left\langle T(L,\omega)\right\rangle\ =  (  2\, \rho_A \, c_A \, \omega)^2
  \left\langle [g_{\omega+}(L,0)]_{11}\ [g_{\omega-}(L,0)]_{22}\right\rangle.
\label{av.T0}
\end{equation}
\  Substituting Eqs.(\ref{gR}), (\ref{gA}) and (\ref{SL0}) into (\ref{av.T0}),
the average transmission rate can be written as
\begin{equation}
\langle T(L,\omega)\rangle\ = \left\langle
\frac{1}{[S(L,0)]_{11}\ [S(L,0)]_{22}} \right\rangle.
\label{av.T1}
\end{equation}
For convenience, we introduce a set of functions
$\left\{A_n(L)\right\}$ which are defined by
\begin{equation}
A_n(x)\ = \ \left\langle  \frac{\{[S(x,0)]_{12}\ [S(x,0)]_{21}\}^n}
{\{[S(x,0)]_{11}\ [S(x,0)]_{22}\}^{n+1}}\right\rangle.
\label{An}
\end{equation}
The averaged transmission rate $\langle T(L,\omega)\rangle$ is equal to
 $A_0(L)$.
{}From Eq.(\ref{S.mat}), we obtain the relation between $S(x,0)$ and
$S(x-\Delta,0)$
\begin{equation}
S(x,0) \ = \ \left(
\begin{array}{cc}
       1+\frac12 |\kappa|^2 & \kappa \\
        \kappa^{\ast} & 1+\frac12 |\kappa|^2
\end{array}\right)
S(x-\Delta,0),
\label{S.m.S}
\end{equation}
where $\kappa$ is defined by Eq.(\ref{kappa.int}) with the integral over the
interval
 $\left[x-\Delta,x\right]$.
 We substitute Eq.(\ref{S.m.S})  into (\ref{An})
  and expand $A_n(x)$ in a power series in $\kappa$  keeping terms up
to $|\kappa|^2$.
\  Thus we get a difference equation for  $A_n(x)$
\begin{eqnarray}
\lefteqn{ A_n(x)  -  A_n(x-\Delta) \ = \ \langle \overline{|\kappa|^2} \rangle}
\nonumber \\
&& \times \left\{ n^2 A_{n-1}(x-\Delta)
 {+}(n+1)^2 A_{n+1}(x-\Delta)
 -\left[n^2+(n+1)^2\right]
    A_n(x-\Delta) \right\}.
\label{def.An}
\end{eqnarray}
In Eq.(\ref{def.An}) we have replaced  the average
$\langle \,  |\kappa|^2 \,\rangle$
with   $ \langle \overline{|\kappa|^2} \rangle$.
\ The average  $ \langle \overline{|\kappa|^2} \rangle $ coincides with
 $ \langle \overline{|\kappa_j|^2} \rangle $ of Eq.(\ref{av.kk3}).
\  Hence, substituting Eq.(\ref{av.kk3}) into (\ref{def.An}) and
taking the limit of $\Delta\rightarrow 0$, we obtain
the following equation for $A_n(t):$
\begin{equation}
\frac{d A_n(t)}{dt}\ = \ n^2 \ A_{n-1}(t) + (n+1)^2\ A_{n+1}(t)
\ - \ \left[n^2\ + \ (n+1)^2\right]\ A_n(t),
\label{dif.An}
\end{equation}
where we have set $x = L$ and changed  the variable to $t$ defined by
$t=L/\ell$.
Next, constructing a  generating function $A(z,t)$ given by
\begin{equation}
A(z,t)\ = \ \sum_{n=0}^\infty \ A_n(t) z^n,
\label{def.A}
\end{equation}
we can derive the partial differential equation satisfied by $A(z,t),$
\begin{equation}
\frac{\partial A}{\partial t} \ = \ z(1-z)^2 \
\frac{\partial^2 A}{\partial z^2} \ +
\ (1-z)(1-3z) \frac{\partial A}{\partial z}\ - \ (1-z)A.
\label{dif.A}
\end{equation}
At $t=0$ or $L=0$,   $A_n=\delta_{n,0}$, since
the $S$-matrix becomes a  unit matrix. Hence, the boundary condition for
the generating function is $A(z,t=0)=1$. The solution of Eq.(\ref{dif.A})
satisfying this boundary condition is\cite{abrikosov}
\begin{equation}
A(z,t)\ = \ \int_0^\infty \ \frac{2\pi \lambda \tanh \pi \lambda}
{\cosh \pi \lambda}
\ \frac{1}{1-z} \ \ F\left(\frac12+i \lambda, \frac12 - i \lambda, -1;
\frac z{1-z} \right) e^{-\left(\frac14+\lambda^2\right)t} d\lambda,
\label{sol.A}
\end{equation}
where $F(\alpha,\beta,\gamma;y) $ is the associated hypergeometric function.
\  Since the ensemble averaged transmission rate $\langle T(L,\omega )\rangle$
is
given by  $A(z=0,t)$, the final expression for $\langle
T(L,\omega )\rangle$ yields
\begin{equation}
\langle T(L,\omega)\rangle\ = \
 \ \int_0^\infty \ \frac{2\pi \lambda \tanh \pi \lambda}{\cosh \pi \lambda}
\exp \left[-\left(\frac14+\lambda^2\right)\ t \right] d\lambda.
\label{av.TF}
\end{equation}
It should be noted that $\langle T \rangle$ is a monotonically decreasing
function of $t$.
We also note that  $\langle T \rangle$ depends on phonon frequency only through
the
phonon mean-free-path $\ell$.
\newpage
\begin{center}
\begin{large}
A. Single-Layer Random SL's
\end{large}
\end{center}
\noindent
We shall now apply the analytical result (\ref{av.TF}) for
$\langle T \rangle$ to real systems  consisting of AlAs-GaAs multilayers.
First, we consider single-layer SL's constructed by randomly stacking   two
kinds of  basic  blocks, one being an AlAs ($A$) single layer
with thickness $d_{AlAs}$ and the other
a GaAs ($B$) single layer with thickness $d_{GaAs}$ [Fig.1(a)].
The interfaces are assumed  to be (100)-plane.
The stiffness constant $\mu$ of our model is the average of the $C_{44}$
moduli for AlAs and GaAs.
This approximation is good since $C_{44} \
(= 5.89 \times 10^{11} \ dyn\, cm^{-2})$ for AlAs is very close to the value
for GaAs ($5.94 \times 10^{11} \ dyn\, cm^{-2}$), but
$\rho_{AlAs}\ (=3.76 \ g \, cm^{-3})\;$ and
$\rho_{GaAs}\ (=5.89 \ g \, cm^{-3})\;$ are quite different.

The ensemble average of the phase factors, $\epsilon_A=\epsilon_{AlAs}$ and
$\epsilon_B=\epsilon_{GaAs}$, are calculated as
\begin{equation}
\epsilon_{AlAs} = \frac{e^{- i a}} {2 - e^{- i a}}
  ,\hspace*{0.5in}
\epsilon_{GaAs} = \frac{e^{- i b}} {2 - e^{- i b}},
\label{SLrandom SL.E}
\end{equation}
where $a \equiv 2\omega d_{AlAs}/c_{AlAs} $,  and
$b \equiv 2\omega d_{GaAs}/c_{GaAs}$, respectively.
The average thicknesses of the segments are also evaluated as
\begin{equation}
\langle D_{AlAs}\rangle=2d_{AlAs},\hspace*{0.5in}
\langle D_{GaAs}\rangle=2d_{GaAs}.
\label{SLrandom SL.D}
\end{equation}
Substituting Eqs.(\ref{SLrandom SL.E}) and (\ref{SLrandom SL.D})
 into (\ref{ell.SL}) and (\ref{ell.b}),
we obtain the expression for the mean-free-path $\ell$.
The dominant part, $\ell_{SL}$, of $\ell$  is
\begin{equation}
\ell_{SL}^{-1}=\frac{4 R^2}{\langle D_{AlAs} \rangle +\langle D_{GaAs} \rangle
}
     \ \  \frac{(1-\cos a)(1-\cos b)}
        {3 - 2 \cos a - 2 \cos b + \cos(a-b)},
\label{ell.SL.Srandom SL}
\end{equation}
where $R = \ R_{AlAs,GaAs} \ =\ (Z_{AlAs} - Z_{GaAs})/(Z_{AlAs} + Z_{GaAs})
= 0.091\ $ for a transverse mode.

To verify the analytical results we have derived, numerical simulations
based on the transfer matrix method are also made.
Figure 2(a) compares $\langle T \rangle$ versus frequency up to $1.3\ $THz
for transverse phonons propagating along the growth direction of
 random SL's with $1000$ blocks.
The thicknesses of the blocks are chosen as $d_{AlAs}=d_{GaAs}=34$\AA, \,
which correspond to twelve mono-atomic layers.
The solid line indicates the analytical result. The open circles are
the numerical data obtained by averaging over 100 random SL's with different
sequences of blocks.
\ The transmission rate of the periodic SL consisting of alternating AlAs and
GaAs blocks is also shown with a dotted line for comparison.

The average transmission $\langle T \rangle$ shows alternating
prominent dips and sharp peaks.
\  The large dips in transmission reflect the localization of phonons
due to interference effects among the scattered waves.
The transmission is perfect or almost perfect at the frequencies satisfying the
resonance conditions  $( \ell_{SL} = \infty )$.    As the sound velocities of
the TA-mode waves are
$c_{GaAs} = 3.32  \times 10^5 \ cm/s \/$ and
$c_{AlAs} = 3.97  \times 10^5 \ cm/s $, the resonance frequencies are
$\nu_{GaAs,n}^{(R)} = n \times 0.489 \ {\rm THz} \/$ for a GaAs single-layer
and
$\nu_{AlAs,n}^{(R)} = n \times 0.583 \ {\rm THz} \/$ for an AlAs single-layer.

We see in Fig.2(a) the slight reductions of $\langle T \rangle$  from unity at
$\nu = \nu_{AlAs,n}^{(R)} \  (n = 1, 2)$. These are due to the scattering of
phonons at both ends of random SL's.
Although $\ell_b^{-1}$ goes to 0 as $L \rightarrow \infty$,
 the  deviation of $\langle T \rangle$ from unity
 survives even for infinitely large systems, because $L/\ell_b$ is a constant
regardless of the system size $L$.

Finally, we compare  $\langle T \rangle$ with the behavior  of the
phonon transmission in  the corresponding periodic SL.
We see that the minima of  $\langle T \rangle$ always occur
in the band gaps of the periodic SL.
{}From Eq.(\ref{ell.SL.Srandom SL}) we can show that the frequencies
at the minima of $\langle T \rangle$  coincide with the
Bragg frequencies\cite{tn90} $\nu_n^{(B)}$ of the corresponding
periodic SL satisfying $\sin \left[(a+b)/2\right] = 0$.
\ This apparent
similarity in the transmission dips between the random and periodic
SL's  is quite interesting.
Of course, the periodic and random SL-systems are very different.
The periodic SL has a finite band width. However, as will be discussed below,
the width of the pass bands in the random SL's where the localization length
is longer than the system size tends to zero for $L \rightarrow \infty$.
\pp
\pp
\begin{center}
\begin{large}
B. Double-Layer Random SL's
\end{large}
\end{center}
\noindent
The second type of random SL's to be studied here are the double-layer random
SL's whose basic building blocks are shown in Fig.1(b).
In contrast to the single-layer random SL's, a characteristic feature
of these SL's is that the interfaces between adjacent blocks never
disappear even if we stack two kinds of basic blocks at random.
Denoting the thicknesses of the constituent layers in the first blocks  by
$d_{1,AlAs}$ and $d_{1,GaAs}$, and those of the second block by
$d_{2,AlAs}$ and $d_{2,GaAs}$,
 the ensemble-averaged phase factors $\epsilon_{AlAs}$ and $\epsilon_{GaAs}$
are given by
\begin{eqnarray}
\epsilon_{AlAs} &= &\frac{ e^{- i a_{1}} +  e^{-i a_2}}2,
\nonumber \\
\epsilon_{GaAs} &= &\frac{ e^{- i b_1} +  e^{-i b_2}}2,
\label{DLRSL.E}
\end{eqnarray}
where $a_{1(2)}=2\omega d_{1(2),AlAs}/c_{AlAs}$ and
$b_{1(2)}=2\omega d_{1(2),GaAs}/c_{GaAs}$.
The average thicknesses of the segments are
\begin{equation}
\langle D_{AlAs} \rangle = \frac{d_{1,AlAs}+d_{2,AlAs}}2,
\hspace*{0.5in}
\langle D_{GaAs} \rangle = \frac{d_{1,GaAs}+d_{2,GaAs}}2.
\label{DLRSL.D}
\end{equation}
Hereafter we consider the case where $d_{1,AlAs} = d_{2,AlAs} \equiv d_{AlAs}$
for simplicity.
For this type of random SL, $\ell_{SL}^{-1}$ becomes
\begin{equation}
\ell_{SL}^{-1} = \frac{R^2} {\langle D_{AlAs} \rangle +\langle D_{GaAs} \rangle
  }
\ \ \frac{(1-\cos a)(1 - \cos (b-c)) }
   {3 - 2 \cos (a+b) - 2 \cos (a+c)  + \cos(b - c)},
\label{ell.DRSL}
\end{equation}
where $a = a_1 = a_2,\ b= b_1$ and $c =b_2$.

 The evaluation of $\langle T \rangle$ for double-layer random SL's is
performed by assuming    100 blocks $(M=100)$
with $d_{AlAs} (=d_{1,AlAs}=d_{2,AlAs})=17{\rm \AA}$
and $d_{1,GaAs}=20{\rm \AA}$ and $d_{2,GaAs}=42{\rm \AA}$.
These pairs of building blocks are the same as those used
by Merlin et al\cite{merlin2}
for the fabrication of quasi-periodic SL's.

Figure 2(b) shows the analytically calculated $\langle T \rangle$
 together with the numerical data.
Again, the agreement between these two calculations is excellent.
The frequency dependence  of $\langle T \rangle$ looks  quite different from
Fig.2(a) for the single-layer SL's  due to the  difference in the structures
of the  building blocks.
In the present case, as can be seen from Eq.(\ref{ell.DRSL}),
the resonances occur for    $a =2 n \pi$ for the AlAs layers
and  $b - c = 2 n \pi$ for the GaAs layers.
The first condition implies that this resonance happens when an
integer-multiple  of the wavelength becomes equal to twice the thickness
$d_{AlAs}$ of an
AlAs layer.  Numerically, the corresponding resonance frequency is given
by  $ \nu_{AlAs,n}^{(R)}=n \times 1.17$  THz.
The second condition means that the thicknesses of   GaAs layers
in the first and second blocks
become effectively identical at the frequencies satisfying this 
condition (numerically $\nu_{GaAs,n}^{(R)} = n \times 0.757$ THz),
implying  that the SL is equivalent to a periodic SL with AlAs layers
of thickness $d_{AlAs}$ and GaAs layers of thickness $d_{1,GaAs}$.
Thus, the phonons can propagate through the random
SL without seeing any of the structural disorder present in it.

Narrow and wide band gaps coexist in the periodic double-layer SL
made with the same unit blocks as the ones used for the random
double-layer SL's.  For these double-layer structures, the location of
the dips in the phonon transmission rate $T$ (for the periodic SL) and
those in $\langle T \rangle$ (for the random SL's) are somewhat different.
\ This difference is much more pronounced when comparing
$T$ and $<T>$ for the single-layer SL's.
\ In the latter, the transmission dips in $<T>$ do not appear at the
frequencies within the narrow gaps present in the $T$ of the periodic SL.
\ We also note that the frequencies at which
the perfect and almost perfect transmissions  occur
coincide with the $\nu^{(R)}$ given above.

\section{LYAPUNOV EXPONENT}

Localized states can be characterized by the localization length $\xi$,
or the Lyapunov exponent $\gamma$, which is the reciprocal of the
 localization length, {\it i.e.}, $\gamma = 1/\xi$.
\  The Lyapunov exponent $\gamma$ is defined by the average of the
logarithmic decrement of the transmission rate as\cite{lfs};
\begin{equation}
\gamma\ = \ - \lim_{L\rightarrow\infty}\frac1{2L}
\left\langle \log T(L,\omega)\right\rangle.
\label{gamma}
\end{equation}
The Lyapunov exponent is subject to the following differential recursion
relation\cite{abrssc}  for an arbitrary function $f(z)$ of
$z (= 1/T )$:
\begin{equation}
\frac{\partial}{\partial t} \langle f\rangle \ =
\ \langle (2 z -1)\frac{\partial f}{\partial z}  \rangle +
\langle z(z-1)\frac{\partial^2 f}{\partial z^2}  \rangle.
\label{d.r.r}
\end{equation}
The derivation of this relation is given in the Appendix.
Putting $f(z)= \log z$  with the initial condition
$\langle f(z) \rangle =0$ at $t=0$ ($ L = 0 $ or $ \ell \rightarrow \infty$),
we find
\begin{equation}
\langle -\log T \rangle \ = \ t.
\label{logT}
\end{equation}
{}From Eqs.(\ref{gamma}) and (\ref{logT}), the Lyapunov exponent $\gamma$  is
 given by
\begin{equation}
\gamma(\omega) \ = \frac1{2\ell_{SL}(\omega)}.
\label{gamma.F}
\end{equation}

Figure 3(a) shows the analytically obtained  Lyapunov exponent
$\gamma \; ( \propto I_s(L\rightarrow \infty) )$
versus frequency $\nu$ together with the numerical data for the single-layer
random SL's.
The numerical data are obtained for the random
SL's with $1500$ blocks for which  the size effect is negligibly small.
 We see that $\gamma$  oscillates regularly
 and its peaks  occur periodically at frequencies
 $\nu_n^{(B)} =n\times0.266 $ THz, which coincide with the Bragg frequencies
of the periodic SL (the height of the peaks is the same,
{\it i.e.},  $1.158\times10^{-4}$\AA$^{-1}$).
The Lyapunov exponent vanishes at the resonance
frequencies $\nu_{j,n }^{(R)} $( $j$ = AlAs or GaAs)
and varies parabolicly in the neighborhood of these frequencies
\begin{equation}
\gamma(\nu) \ =  \frac{8 \pi^2 R^2}{d_{AlAs}+d_{GaAs}} \,
              \frac{d_j^2}{c_{j}^2} \left(\nu-\nu_{j,n }^{(R)} \right)^2.
\label{gamma.GaAs}
\end{equation}
\noindent

In a finite system, phonons whose localization length is longer than
the length of the system are regarded as extended.
Thus, the critical localization
length $\xi_{cr} = \gamma_{cr}^{-1} \equiv L \ \ ( \ell_{cr} = L/2 )$
discriminates between localized and delocalized  states.
We can define the phonon pass band
as a frequency region where the Lyapunov exponent $\gamma$  is smaller than
the critical value $\gamma_{cr}$. The width of the pass bands $\Delta\nu$ is
estimated from Eq. (\ref{gamma.GaAs})  as
\begin{equation}
\Delta\nu =  \frac1{\pi R} \, \sqrt{
     \frac{d_{AlAs}+d_{GaAs}}{ 2L}} \, \frac{c_j}{d_j } .
\label{pass.band}
\end{equation}
Thus, the pass band width vanishes in an infinitely large SL as noted above.
 The average transmission  at the band edges is readily
estimated from Eq.(\ref{av.TF}) by putting $t = t_{cr}=L/\ell_{cr} =2$.
We find $\langle T \rangle = T_{cr} = 0.26$.
Therefore,  $\langle T \rangle$ is greater
than $0.26$ in the pass bands.

Figure 3(b) shows the Lyapunov exponent $\gamma $ for the double-layer SL's,
which exhibits quite a different frequency dependence from $\gamma$ in the
single-layer SL's.
In addition, the magnitude is significantly larger than $\gamma$ in
the single-layer random SL's.   These differences are attributed to the
internal structure of the constituent blocks.  In particular, the interfaces
between any adjacent blocks are recognized by phonons, leading to a much
shorter phonon mean-free-path as compared to single-layer SL's
with the same number of unit blocks.  This also leads to conspicuous dips in
$\langle T \rangle$ in the double-layer SL's even when the number of blocks
involved in a random SL is small (see Fig.2(b)).

The parabolic behavior of the Lyapunov exponent $\gamma$ near the resonance
frequencies and the size of a pass band width can be derived in
a similar  way as for the single-layer SL's but these calculations
are not presented here.

\section{UNIVERSALITY IN  TRANSMISSION  \newline
\hspace{2.0in}    \  FLUCTUATIONS}
The phonons in a pass band  are delocalized and their transmission rates
are expected to fluctuate due to interference among multiply scattered phonons,
just like the electronic conductance fluctuations in mesoscopic
transport\cite{ALS,nscf}.
 The standard deviation of the transmission $\Delta T$, defined by
\begin{equation}
\Delta T \ = \ \sqrt{ \langle T^2 \rangle - \langle T \rangle^2},
\label{dT}
\end{equation}
can be readily calculated by using the formula derived from Eq.(\ref{d.r.r}),
\begin{equation}
\langle T^2 \rangle \ =  -\ \frac d{dt} \langle T \rangle.
\label{TT}
\end{equation}
Thus, the transmission  fluctuation $\Delta T$  is also a function of
   $t=L/\ell$, and $\Delta T$
does not depend on the details of $\ell$, or the structure of the random SL's.
\ This means that  the relation between $\Delta T$ versus $\langle T \rangle$
should be universal, independent of the type of SL.

Figure 4(a) shows $\Delta T$ versus $\langle T \rangle$
together with the numerical data for the single- and double-layer random SL's.
Indeed, no difference can be seen in the numerical results of $\Delta T$
versus $\langle T \rangle$ 
and the analytical result coincides remarkably well with the numerical data.
\ For localized phonons ($ \langle T \rangle \sim 0 $), the fluctuations
appear to be small and they increase with increasing $\langle T \rangle$.
The maximum of $\Delta T$ is attained in the pass band at
$\langle T \rangle \simeq 0.4$ where the mean-free-path becomes
comparable to the system size, or $L / \ell = 1.27 $ .

Although the transmission fluctuations are small for localized phonons, the
transmission rate cannot properly be averaged for these
phonons.  We plot the relative fluctuations of transmission in Fig.4(b).
\  The relative fluctuations diverge  at $\langle T \rangle=0$ and
decrease  with  increasing  $\langle T \rangle$.
 At the boundary between the localized and extended states,
 the relative fluctuations become almost unity and reduce to zero at
$\langle T \rangle =1$.  The relative fluctuations always exceed unity for
the localized phonons, therefore, the transmission rate is not  well-defined in
the localized regime.

\ It is known that, in a one-dimensional metal, the conductance or the
resistance has the same properties as the phonon transmission rate discussed
here. Anderson et al.\cite{andrsn} applied a method of scaling transformation
and found that the logarithm of the total resistance obeys a normal
distribution for large sample lengths.

In order to see the behavior of relative fluctuations of   $ \log T $,
we evaluate the ratio of the
standard deviation of $\log T$ to $\langle \log T \rangle$.
With the use of Eq.(\ref{d.r.r}), the standard deviation of $\log T$ is given
by
\begin{equation}
\Delta\log T \ = \ \left[ 2 \int_0^t
             \left\{ 1-\langle T \rangle \right\} dt \right]^{1/2}.
\label{DlogT}
\end{equation}
Figure 5 shows the analytical and numerical results for the relative
fluctuation of $\log T$.
Our analytical result indicates that the relative fluctuations
become unity at $\langle \log T \rangle = 0$  and decreases
monotonically with increasing  $t$.
Thus, we conclude  that $\log T$, in contrast to $T$, is a well-defined
quantity
 for localized phonons.

The peculiar behavior of $\Delta T$ versus $\langle T \rangle$
can be explained in terms of  the fluctuations
of the Lyapunov exponent or the localization length plotted in Fig.5.
  Here we note that $\langle \,\log T \, \rangle$ is proportional to the
Lyapunov exponent $\gamma$.
The relative fluctuation of $\log T$ is, hence,  equivalent to
$\Delta \gamma/\gamma$.  For highly extended phonons with
$\langle T \rangle \simeq 1$ where $\gamma$ is close to 0,
$\Delta \gamma$ has the same magnitude as
$\gamma$ ( $ \Delta \gamma / \,\gamma \simeq 1 \  {\rm from \; Fig.~5} $ ),
{\it i.e.}, $\; \Delta \gamma \sim 0$.
This is the reason why  $\Delta T$ is small in the region
$\langle T \rangle \simeq 1$.   The relative fluctuation
$\Delta \gamma/\gamma$ decreases with increasing $\gamma \; ( \propto t )$. The
relative fluctuation of the localization length,
$\Delta \xi/\xi$, is equal to $\Delta \gamma/\gamma$ and hence,
$\Delta \xi$ is always smaller than $\xi$. In the extended region,
the SL size $L$ is shorter than $\xi$ but might become comparable
to $\Delta \xi$ because $ \Delta \xi  \,  \stackrel{<}{\sim} \, \xi$.
In this region, the transmission should be largely affected
by fluctuations of the localization length or Lyapunov exponent.
The fluctuation $\Delta \xi$ is much shorter than the SL size $L$ in the
localized regime since $\Delta\xi < \xi \ll L$. Therefore,
the fluctuation $\Delta T$ becomes quite small and vanishes at
$\langle \, T \, \rangle \simeq 0$.  However, we should keep in mind that
the fluctuation $\Delta \xi$ still affects the transmission in this regime
because $\Delta T /\, \langle T \rangle$ diverges at
$\langle \, T \, \rangle \simeq 0$  as discussed above.

\pp
\section{ LOG-NORMAL DISTRIBUTION OF \newline
   TRANSMISSION}
In this section, we study the probability distribution of $T$ in  the random
SL's.  At the resonance frequencies, the transmission rate
is unity and its distribution function should have a peak like $\delta(T-1)$.
\  From the analogy with the electronic localization problem, the probability
distribution for localized phonons is expected to be Gaussian
with respect to $\log T^{-1}$.  We are interested in the transition of the
distribution with decreasing average transmission.
We here introduce a probability density $W(z,t)$ with which one can express
the average of an arbitrary function $f(z)$ of $z=1/T$ as
\begin{equation}
\langle f \rangle \ = \ \int_1^\infty f(z) W(z,t) dz.
\label{av.f}
\end{equation}
Substituting Eq.(\ref{av.f}) into (\ref{d.r.r}) and integrating by parts,
we obtain the Fokker-Planck equation for $W(z,t)$:
\begin{equation}
\frac{\partial W}{\partial t} \ = \ - \frac{\partial}{\partial z}
\left[(2z-1)W\right]+
\frac{\partial^2}{\partial z^2} \left[z(z-1)W\right].
\label{eq.of.W}
\end{equation}
Here the initial condition is $W(z,t=0)=\delta(z-1)$.
Equation (\ref{eq.of.W}) was studied by Abrikosov and solved\cite{abrssc}
with the associated hypergeometric functions as
\begin{equation}
W(z,t)\ = \ \frac2{\sqrt{\pi t^3}} \int_{x_0}^\infty \ \frac x {\sqrt{\cosh^2
x-z}}
\exp \left[- \left(t/4+x^2/t\right) \right] dx,
\label{W}
\end{equation}
where $x_0= \cosh^{-1}\sqrt{z}$.

Figure  6(a)  shows the statistical distribution of $\log T$ for the extended
phonons ($t = 0.47$  and  $1.18$, and the corresponding $\langle T \rangle$ are
$ 0.67$  and $0.42$, respectively)  and the critical states ($t = 2$, and
correspondingly
 $\langle T \rangle = T_{cr} = 0.26$) together with the numerical data.
The similar plot for the localized phonons ($t = 8.22$, and correspondingly
 $\langle T \rangle =  0.02$) is  given in Fig.6(b).
The ordinate has been taken as $W(z, t) \ z$ because
\begin{equation}
W(z,t) dz =   W(z,t) \ \frac{dz}{d \log T} \  d \log T = - W(z,t) \; z \;
d(\log T).
\end{equation}
The distribution for the extended phonons has a peak at the
origin $\log T=0$, suggesting a  Poisson distribution.
The  peak height is lowered with decreasing $\langle T \rangle$.
\  Simultaneously, the FWHM increases and the shape of distribution
curves becomes convex upward, but the position of the peak stays unchanged
($ \log T = 0$).
For the strongly localized states, {\it i.e.}, $\langle T \rangle \ll T_{cr}$,
a Gaussian distribution is attained. The distribution $\; W(z,t) \ z\ $
for the localized states is analytically derived from Eq.(\ref{W})
as\cite{abrssc}
\begin{equation}
W(z,t) z \simeq \frac1{2 \sqrt{\pi t}} \exp \left[ - \frac{(\log T + t)^2}{4 t}
\right],
\ \  \  t \gg 1,
\label{gauss}
\end{equation}
and the peak is located at a finite value of $\log T$, {\it i.e.},  $\log T = -
 t $.
The solid curve of Fig.6(b) for $t = 8.22$ agrees well with Eq.(\ref{gauss}).
The FWHM obtained from Eq.(\ref{gauss}) is $\sqrt{2t}$.
{}From these quantities,  the relative fluctuations of $\log T$ for the
strongly
localized states can be calculated again as $\sqrt{2 t}/t =\sqrt{2/t}$,
which coincides with the asymptotic formula (valid at $ t \gg 1$)
derived analytically from Eq.(\ref{DlogT}) (see Fig.5).

\ To conclude,  the transmission rate in the localized regime obeys  the
log-normal distribution and a Poisson distribution applies to highly extended
phonon states.

\section{ SUMMARY AND CONCLUSION}

\indent
Based on the Green's function method, we have studied the transmission of
phonons propagating along the growth direction of random SL's.

Taking  account of the phase shift of phonons associated with the
forward scattering due to the mass density fluctuations, we have derived
an explicit expression for the average transmission $\langle T \rangle$
as a function of the scaling parameter $t=L/\ell$, {\em i.e.}, the size of the
SL divided by the mean-free-path of phonons limited by backscattering.
The mean-free-path  $\ell$ reflects the structural
properties of the SL and is expressed by the average of the squared
SL structure factor, $\langle |S_{SL}(L,\omega)|^2 \rangle$.
We can readily evaluate the average transmission $\langle T \rangle$ for a
random SL, by calculating analytically the mean-free-path $\ell$.
We applied the analytical expression for the average transmission
$\langle T \rangle$ to two types of SL's with different unit-block
structures, {\em i.e.}, single- and double-layer random SL's. The
average transmission rates show dips and peaks due to coherent
backscattering and resonances of phonons  depending on the structure of
the SL. The coincidence of the analytical results
with the numerical data based on the transfer matrix method verifies
the accuracy of our theory.
The universality  relation of $\Delta T$ versus $\langle T \rangle$ has
also been established explicitly  for both types of SL's.
Finally, we have examined the statistical distribution of the phonon
transmission and have confirmed a log-normal distribution for the localized
phonons.

The most important result we have found in the present work is that the
phonon-mean-free-path (and eventually  $\langle T \rangle$) can be calculated
from  $\langle |S_{SL}(L,\omega)|^2 \rangle$.  This enables us to study
quantitatively the average transmission  for a given set of SL's with
simple calculations of the mean-free-path based on the prescribed
statistical rules for the building blocks.

The existence of resonances, in the SL system we have considered, yields an
interesting structure in the phonon transmission spectrum, such as the
alternating regions, in the frequency domain, of localized and extended states.
\ By modulating the thickness of the constituent layers, we can readily
produce resonances in the sub-THz frequency region, which are accessible
experimentally. In pure semiconducting materials, the phonon mean free path
limited by the bulk elastic scattering caused by mass defects
(foreign and isotopic impurities etc.) is 0.1 to 1 cm, or much
longer at frequencies below 1THz.  Also, inelastic scattering of phonons is
much weaker at low temperatures.  Hence,
semiconductor hetero-structure random SL's
would permit the observation of the effects of
high-frequency phonon multiple-scattering induced by the mass-density
fluctuations in the layered structures.

Through this work we have assumed the same stiffness constants for the SL
constituent materials.  However, this does not limit the applicability of
our results because the fluctuations  in the stiffness constants are
effectively incorporated into  the mass-density fluctuations in the present
formulation.  Thus, our results are readily applicable to SL's with various
combinations of materials, and also to the study of propagation of third-sound
waves in HeII films adsorbed on disordered substrates\cite{third}.
\begin{center}
 \begin{large}
Acknowledgments
\end{large}
\end{center}
\pp
The authors would like to thank T. Sakuma and R. Richardson  for useful
comments on the manuscript.  This work is supported in part by the Special
Grant-in-Aid for Promotion of Education and Science in Hokkaido University
Provided by the Ministry of Education, Science and Culture of Japan,
a Grant-in-Aid for Scientific Research from the Ministry of Education,
Science, and Culture of Japan (Grant No. 03550001), the Suhara Memorial
Foundation, and the Iketani Science and Technology Foundation.  FN
acknowledges partial support from the
NSF grant DMR-90-01502, a GE Fellowship, and SUN microsystems.

\newpage
\begin{center}
 \begin{large}
APPENDIX:\\
\pp
Derivation of Differential Recursion Relation
\end{large}
\end{center}
\pp

Consider an arbitrary function $f(z)$ expanded as
$$ f(z)   =   \sum_n a_n z^n, \eqno{(A.1)} $$
where $z=1/T(L,\omega)=[S(L,0)]_{11} [S(L,0)]_{22}$.
Substituting Eq.(\ref{S.m.S}) into (A.1) and expanding the function with
respect to
$\kappa$ and $\kappa^\ast$, an ensemble average of a term $z^n$
is given up to $O(|\kappa|^2)$ by
$$\langle z_1^n \rangle   =  \langle z_0^n \rangle +
  \langle \overline{|\kappa|^2} \rangle \{ n  \langle z_0^{n-1} ( 2
z_0-1)\rangle
+ n(n-1)  \langle z_0^{n-1} (  z_0 -1)\rangle \}, \eqno{(A.2)}$$
where $z_1=z(L)$ and $z_0=z(L-\Delta)$.
Taking the limit of $\Delta \rightarrow 0$ and dividing
 (A.2) by $\langle \overline{|\kappa|^2} \rangle$,  we obtain  the following
differential  equation for the $\langle z^n \rangle$ term
$$
\frac{\partial}{\partial t} \langle z^n \rangle  =
  \langle \frac{\partial z^n}{\partial z} (2 z -1) \rangle +
      \langle \frac{\partial^2 z^n}{\partial z^2} z(z -1) \rangle. \eqno{(A.3)}
$$
Thus, the differential recursion
relation for an arbitrary function  $f(z)$
is derived as
$$
\frac{\partial}{\partial t} \langle f(z) \rangle   =
  \langle \frac{\partial f(z)}{\partial z} (2 z -1) \rangle +
      \langle \frac{\partial^2 f(z)}{\partial z^2} z(z -1) \rangle.
\eqno{(A.4)}
$$
This equation was originally derived by Abrikosov\cite{abrssc}
 for the study of the electrical conductivity in disordered systems.

\newpage
{\Large \bf Figure captions}
\vskip 0.2in

\pp
Fig.~1
\  Schematic representations of the building blocks for (a) single-layer SL's
and
(b) double-layer SL's made of two kinds of materials $A$ and $B$.
In the numerical calculations AlAs and GaAs are assumed for the
materials A and B, respectively.

Fig.~2  \
 Average phonon transmission rate, $\langle T \rangle$, versus frequency,
$\nu$,
for (a) single-layer random SL's with 1000 building blocks ($N = 1000$) and
(b) double-layer random SL's with 100 building blocks ($ N = 199$).
The solid lines show the analytical results and the open circles are the
numerical data calculated from  the transfer matrix method. Each data
point is obtained by averaging over $100$
realizations of SL randomness. The parameters used  are:
$d_{AlAs} = d_{GaAs} = 34$\AA \, for the
single-layer random SL's used in (a); and $d_{1,AlAs} =d_{2,AlAs} =d_{AlAs} =
17$\AA, $d_{1,GaAs} =  20$\AA \, and $ d_{2,GaAs} =  42$\AA \,
for the double-layer random SL's used in (b). The mass densities are
$\rho_{AlAs} = 3.76 \,  g\, cm^{-3}  \ {\rm and } \
\rho_{GaAs}  = 5.36 \,  g\, cm^{-3}; \ {\rm and } \
\mu          = 5.915\times 10^{11} \, dyn\,cm^{-2}.$
The slight shifts of resonance frequencies
($\nu_{GaAs,n}^{(R)}$ and $\nu_{AlAs,n}^{(R)}$) from those in Ref.3
are attributed to the average stiffness constant $\mu$ used in the present
model.  The dotted line shows the transmission rate of a periodic SL with
the same constituent blocks.

\pp \
Fig.~3  \ Lyapunov exponent  $\gamma \;$  (the reciprocal of the localization
length $\xi$) versus  frequency, $\nu$, for (a) the single-layer random SL's
and
(b) the double-layer random SL's.
The solid lines show the analytical results and the open
circles are the numerical data.
The numerical results are obtained  for (a) the single-layer
random SL's with $N=1500$ and (b) the double-layer
 random SL's with $N=199$, respectively. The data are
averaged  over $100$ possible realizations of SL randomness.

\pp
Fig.~4 \
(a) Standard deviation of the phonon transmission rate,
$\Delta T$, versus $\langle T \rangle$.  The continuous line denotes the
analytical result and the open squares and  circles are the numerical data
obtained for (a) single- (open squares) and (b) double- (open circles)
layer random SL's, respectively. Each data point is obtained by averaging
over 100 different SL's.  (b) The relative transmission fluctuation
$\Delta T/\langle T \rangle$ versus the average, $\langle T \rangle$.
 The solid line is the analytical result and open
circles are numerical data for the two types of random SL's considered.
\ $T_{cr} = 0.26 \  ( L = 2 \ell) \;$ indicates the boundary
between the extended and localized regimes for phonons.
\pp
Fig.~5 \
Relative fluctuation of $  \log T$ versus $- \langle   \log T \rangle = t$.
\ The solid lines show the analytical results and the open circles
are the numerical data for the two types of random SL considered.
\ The dashed line is the asymptotic formula $(t/2)^{-1/2}$ derived
analytically for $t \gg 1$.

\pp
Fig.~6 \
Probability distribution of $\; W(z, t) \; z \ $ versus $\; - \log T \;$
togetherwith the numerical data for
(a) extended phonons ($\langle T \rangle > T_{cr} = 0.26 $)
with $\langle T \rangle =0.67$ \ (solid  line   and  circles),
$0.42$ \ (dashed line and squares) and critical states with
 $\langle T \rangle = T_{cr}$, {\it i.e.}, $L = \xi_{cr}$\
(dot-dashed line and triangles), and for
(b) localized phonons with $\langle T \rangle =0.02$ \ (solid line and
squares).
The arrows indicate the positions of $t = \langle - \log T \rangle$.
Note that $z = 1/T$.
The numerical data are obtained for an ensemble of 1000 single-layer SL's
by calculating $T$ at a single frequency giving the chosen value of
$\langle T \rangle$.

\end{document}